\documentclass[mypaper,8pt,twoside]{CoAst}
\usepackage{epsf,graphicx,fancyhdr,sfmath}
\renewcommand{\chaptermark}[1]{\markboth{#1}{}}

\newcommand{\teff}{\ensuremath{T_{eff}}}             
\newcommand{\logg}{\ensuremath{\log g}}                     

\newcommand{\kopf}{\small\itshape Comm.\ in Asteroseismology, N$^{\textsf{\underline{o}}}$ 159, 2009\\
Proceedings of the JENAM 2008 Symposium N$^{\textsf{\underline{o}}}$~4:
Asteroseismology and Stellar Evolution}

\newcommand{\Authors}[1]{\begin{center}\normalsize\bf\sf #1 \end{center}}

\renewcommand{\author}[1]{\begin{center}\normalsize\bf\sf #1 \end{center}}
\newcommand{\Address}[1]{\begin{center}\small\sf #1 \end{center}}

\newcommand{\Objects}[1]{{\vspace{3mm}\small \noindent Individual Objects: }\small\sf \hangindent=27truemm \hangafter=1 #1 }

\renewenvironment{abstract}{\section*{Abstract}\normalsize\sf}{}
\newcommand{\References}[1]{\begin{flushleft}{\large References\\}\vspace*{2mm}\small #1 \end{flushleft}}

\newcommand{\chapterCoAst}[2]{\chapter[\sf\normalsize #1\\ \footnotesize \hspace*{5mm}by #2 \sf\normalsize][]{#1\\}\rhead[\fancyplain{}{\sf\footnotesize \center{#1}}]{\fancyplain{}{\sffamily\thepage}}\lhead[\fancyplain{\kopf}{\sffamily\thepage}]{\fancyplain{\kopf}{\sf\footnotesize \center{#2}}}}

\newcommand{\figureCoAst}[5]{\begin{figure}[#4]
\centering
\includegraphics*[#5]{#1}
\caption{#2}
\label{#3}
\end{figure}}

\renewcommand{\chaptername}{}

\newcommand{\acknowledgments}[1]{\vspace*{5mm}\noindent  \textbf{Acknowledgments.} #1}

\def\rfr{\smallskip\par\noindent\hangindent=7truemm\hangafter=1}

\newcommand{\lsim}{\raisebox{-0.6ex}{$\stackrel{{\displaystyle<}}{\sim}$}}

\newcommand{\pdot}{$\dot{\rm P}$}

\begin{document}
\sf

\chapterCoAst{Search for sdB/WD pulsators in the Kepler FOV}
{R.\,Silvotti, G.\,Handler, S.\,Schuh, B.\,Castanheira, and H. Kjeldsen} 
\Authors{R.\,Silvotti$^{1}$, G.\,Handler$^{2}$, S.\,Schuh$^{3}$,
B.\,Castanheira$^{2}$, and H. Kjeldsen$^{4}$} 
\Address{
$^1$ INAF -- Osservatorio Astronomico di Capodimonte,
via Moiariello~16, 80131~Napoli, Italy\\
$^2$ Institut f\"ur Astronomie, Universit\"at Wien, T\"urkenschanzstrasse~17, 
1180 Vienna, Austria\\
$^3$ Institut f\"ur Astrophysik, Universit\"at G\"ottingen,
Friedrich-Hund-Platz~1, 37077~G\"ottingen, Germany\\
$^4$ Dep. of Physics and Astronomy, Aarhus University, Ny Munkegade, 
DK-8000 Aarhus C, Denmark} 

\noindent
\begin{abstract}
In this article we present the preliminary results of an observational search
for subdwarf B and white dwarf pulsators in the Kepler field of view (FOV), 
performed using the DOLORES camera attached to the 3.6~m 
{\it Telescopio Nazionale Galileo} (TNG).
\end{abstract}

\Objects{KIC10\_05807616, KIC10\_02020175}

\section*{Introduction}

The Kepler satellite will be launched in March 2009 and will observe
a $\sim$105 square degree field for 4 years with the primary goal of finding
new exoplanets using the transit method.
Kepler's secondary goal is asteroseismology: the objective is to 
characterize stars hosting planets, and also to study in detail 
a few thousands other oscillating stars.
Among the seismic targets, up to 512 stars can be observed in short cadence
with a sampling time of 1 minute (for all the other targets the cadence will 
be 30 min), allowing the study of short period pulsators, including
hot subdwarfs B (sdBs) and white dwarfs (WDs).
Thanks to its exceptional photometric accuracy and duty cycle ($\approx$95\%,
see Christensen-Dalsgaard et al. 2006 for more details), Kepler can produce
numerous exciting results on these stars:
1) detect low-amplitude (\lsim100~ppm) and high-degree ($l$$>$2) modes, not 
visible from the ground.
2) Measure stellar global parameters with unprecedented accuracy (mass, 
rotation, H/He layer thickness, \teff, \logg).
3) Improve our understanding of the physics of these stars (differential rotation; 
core C/O ratio and equation of state, neutrino cooling and crystallization in 
WDs).
4) Study amplitude variations and nonlinear effects.
5) Through the O--C diagram, measure \pdot, determine the evolutionary 
status of the star and search for low-mass companions (BDs/planets) 
with masses down to $\approx$10$^{-1}$M$_{Jup}$ (see the recent example of 
V391~Peg~b, Silvotti et al. 2007).

\section*{Observations and preliminary results}

The 24 targets were selected from the KIC10 (Kepler Input Catalogue version 10,
used internally by the Kepler team to select targets) through their {\it g-i} 
SLOAN colour.
For most of the targets proper motions were available from the USNO catalogue 
allowing to refine the selection using a reduced proper motion diagram.

The time-series photometry was performed during a single run at the 3.6~m TNG
in August 2008.
Each target was observed for 1 to 2 hours with the SLOAN {\it g} filter, 
with exposure times between 1 and 10~s, depending on the magnitude.
Details on the selection criteria and data reduction will be given in a 
forthcoming paper (Silvotti et al. in preparation).

A new variable star was found from a preliminary analysis of our data. 
It is probably a cataclysmic variable and we will present it in a future 
article. 
However, our data did not revealed any star with a clear signature of 
intrinsic pulsations.
The typical upper limits that we have obtained for the pulsation amplitude are 
between 1 and 2 mma (1\,mma = milli-modulation-amplitude = 1000\,ppm).
Nevertheless, for two targets having colours compatible with sdB stars, 
the light curve or the amplitude spectrum suggest possible periodicities.
The upper panel target in Figure~\ref{fig:lc_dft_2} shows a periodicity near 
1 hour, compatible with a slow ({\it g}-mode) sdB pulsator.
The lower panel target has a peak higher than 3 times 
the local noise (S/N$>$3) at about 125~s, which would correspond to a rapid 
({\it p}-mode) sdB pulsator.

New observations of these two stars done in October 2008, and now under 
reduction, will help to clarify whether they oscillate or not.
Just before submitting this article, we have been informed that 
all the 24 targets observed at the TNG have been included in a list of 
{\it survey targets}
that should be observed by Kepler in the first months of the mission to verify 
their pulsational stability with a much lower detection threshold.

\figureCoAst{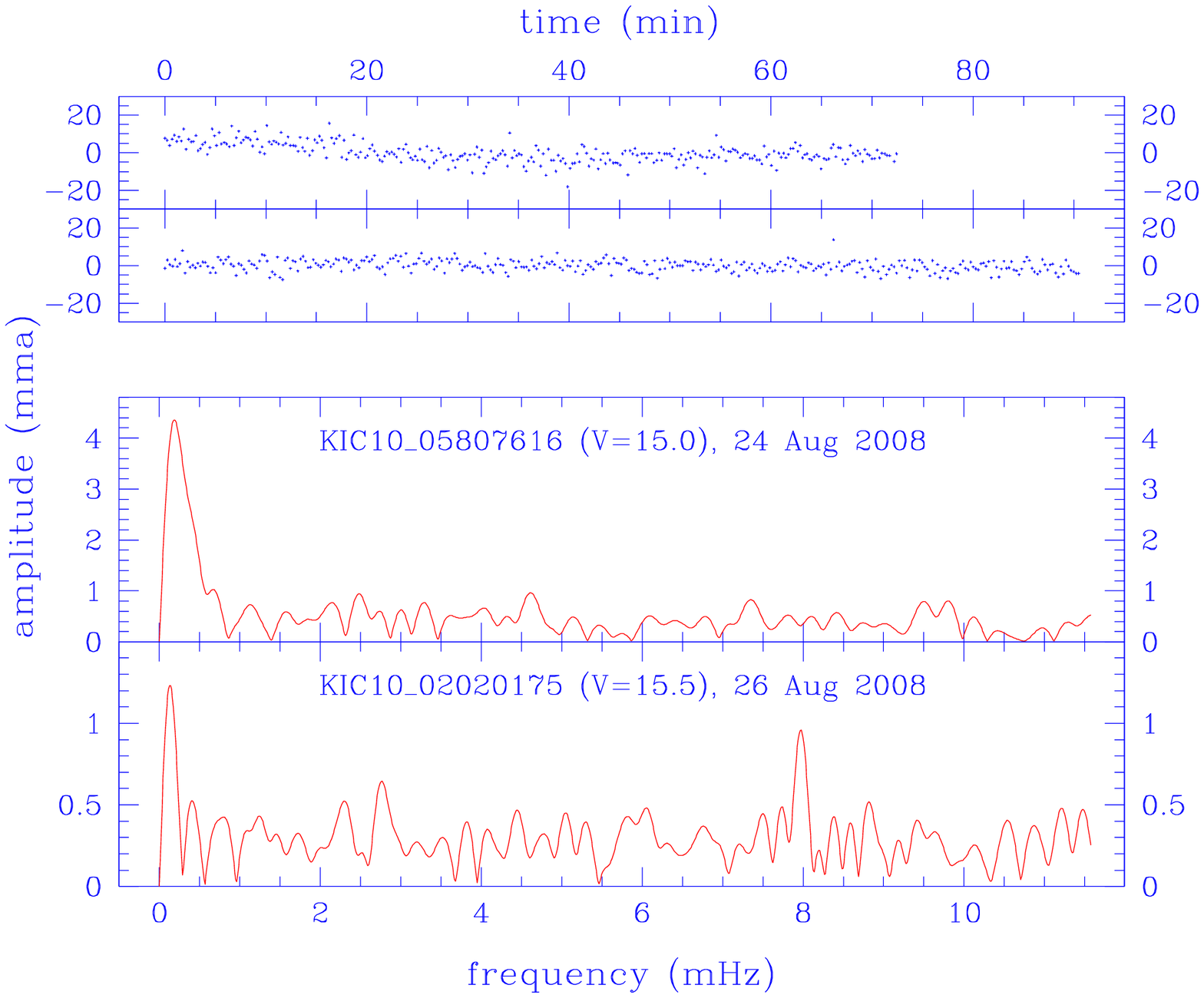}
{Light~curve~and~amplitude~spectrum~of~the~two~best~pulsator~candidates 
(see~text~for~more~details).}
{fig:lc_dft_2}{t}{clip,angle=0,width=0.8\textwidth}

\acknowledgments{
The authors are very grateful to Alfio Bonanno and Silvio Leccia for having
given part of their six TNG nights to this programme, following technical 
problems with the SARG instrument.
Silvio Leccia has been also observing during the first night. 
RS wishes to thank the TNG technical team, in particular Gloria Andreuzzi,
Antonio Magazzu and Luca Difabrizio, who did an excellent job during the 7 
nights of service-mode observations.}

\References{
\rfr Christensen-Dalsgaard, J., Arentoft, T., Brown, T. M., et al. 2006, 
Comm. in Asteroseis. 150, 350
\rfr Silvotti, R., Schuh, S., Janulis, R., et~al. 2007, Nature, 449, 189}

\end{document}